\documentclass[epj]{svjour}

\pdfoutput=1
\usepackage{amsmath}
\usepackage{graphicx}
\usepackage[pdftex,dvips]{epsfig,graphicx}

\newcommand{\be}{\begin{equation}}
\newcommand{\ee}{\end{equation}}
\newcommand{\bma}{\begin{displaymath}}
\newcommand{\ema}{\end{displaymath}}

\begin{document}

\title{Electronic shell and supershell structure in graphene flakes}

\author{M. Manninen, H. P. Heiskanen, and J. Akola}

\institute{NanoScience Center, Department of Physics,
FI-40014 University of Jyv\"askyl\"a, Finland}

\date{\today}

\abstract{
We use a simple tight-binding (TB) model to study electronic properties 
of free graphene flakes. Valence electrons of triangular graphene flakes 
show a shell and supershell structure which follows an analytical 
expression derived from the solution of the wave equation for triangular 
cavity. However, the solution has different selection rules for triangles 
with armchair and zigzag edges, and roughly 40000 atoms are needed to see 
clearly the first supershell oscillation. In the case of spherical flakes, 
the edge states of the zigzag regions dominate the shell structure which 
is thus sensitive to the flake diameter and center. A potential well that 
is made with external gates cannot have true bound states in graphene due 
to the zero energy band gap. However, it can cause strong resonances in 
the conduction band.
\PACS{
      {73.21.La}{Quantum dots} \and
      {81.05.Uw}{Carbon, diamond, graphite} \and 
      {61.48.De}{Structure of carbon nanotubes, boron nanotubes, and closely 
		related graphitelike systems}\and 
      {81.05.Uw}{Carbon, diamond, graphite} 
     } 
} 

\maketitle

\section{Introduction}

Electrons confined in a finite cluster of atoms with spherical symmetry 
exhibit a shell structure\cite{martins1981,deheer1993}. In large enough 
clusters, the shells representing different classical periodic orbits
can interfere forming a supershell structure\cite{balian1970,nishioka1990}
that has been observed in large alkali metal clusters\cite{pedersen1991}.
The supershell structure is especially visible in a two-dimensional
triangular cavity which has only two classical periodic orbits\cite{brack1997}.
The triangular cavity is interesting also due to the fact that the 
Schr\"odinger equation and the wave equation are exactly solvable in that
system\cite{lame1852,krishnamurthy1982,doncheski2003}, and it has been shown 
that triangular shapes are preferred in two-dimensional nearly free electron 
systems\cite{reimann1997,kolehmainen1997,reimann1998}.

Recently, experiments have shown that single layer graphene flakes can be 
prepared on inert surfaces where the graphene-surface interaction is weak
\cite{berger2004,novoselov2004,berger2006,novoselov2007,geim2007,li2007}. 
Since the manipulation of graphene on different substrates is still a fast 
developing area, it is not out of question that graphene flakes with 
accurate shape and size can be eventually processed on a substrate where the 
interaction is so weak that it does not affect the graphene electronic 
levels close to the Fermi point. Hence, we study ideal free graphene flakes 
neglecting the interaction with the substrate. The experiments have 
inspired a wealth of theoretical studies of graphene
\cite{alicea2005,gusynin2005,tworzydlo2006,gusynin2006,zhou2006,yamamoto2006,nomura
2007,son2007,areshkin2007,fernandez2007,chen2008,castro2008,akhmerov2008,zhang2008},
but ackording to our knowledge the shell and supershell structure of large
graphene flakes has not been addressed except in our recent work
\cite{akola2008}.

Electronic structure calculations based on the density functional theory 
(DFT) have shown that the energy levels close to the Fermi level, which 
consists of discrete points in graphene, are determined by the $p$ electrons 
perpendicular to the graphene plane (for a review see \cite{castroneto2008}).
A simple tight-binding (TB) model with only one electron per site and only 
the nearest-neighbour hopping describes well the electronic structure close 
to the Fermi points as suggested by Wallace already in 1947
\cite{wallace1947}. The TB hamiltonian used is then the simple H\"uckel 
model
\begin{equation}
H_{ij}=\left\{\begin{array}{rl}
-t, & {\rm if} \quad i,j \quad{\rm nearest \quad neighbours}\\
0, & {\rm otherwise},\\
\end{array}\right. 
\end{equation}
where the hopping parameter $t$ (resonance integral) determines the width 
of the bands and the on-site energy is chosen to be $\epsilon_F=0$. 
We present our results in units $t=1$ which in real graphene corresponds to 
$\sim$2.6 eV. In reality, the flake edges are either passivated (e.g. with 
hydrogen) or reconstructed in order to remove dangling bonds. The 
passivation is not expected to affect the perpendicular $p$-states, and we 
can simply neglect the existence of such atoms. This has been also validated 
by our recent DFT calculations \cite{akola2008}. At the bottom of the 
valence band the TB model results in free-electron-like states with nearly 
constant density of states (DOS). This allows us to compare these 
``normal'' free-electron states with those of the ``massless electrons'' at 
the bottom of the conduction band calculated for the exactly same geometry.

The paper is organized as follows: In Section 2 we discuss the shell and 
supershell structure in triangular graphene flakes, in Section 3 we show 
how the edge geometry dominates the shell structure in circular flakes, and 
in Section 4 we describe quantum dots that have been made with external 
potentials in an infinite graphene sheet. Section 5 gives the conclusions.

\section{Shell structure of triangular graphene flakes}

\begin{figure}[h]
\includegraphics[width=\columnwidth]{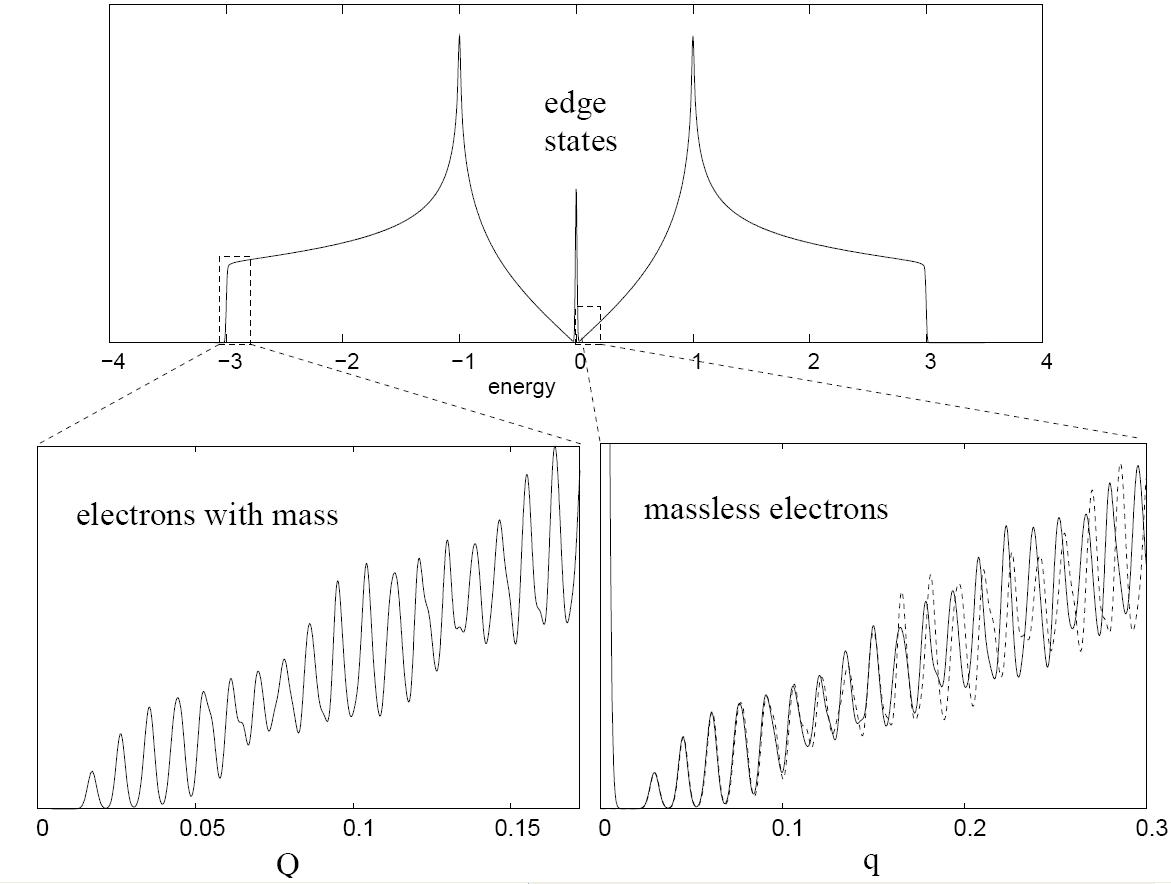}
\caption{Upper panel: 
TB-DOS ($p_z$ electrons) of a graphene triangle with 44097 atoms and zigzag 
edges. The discrete energy levels have been smoothened with Gaussians. The 
peak at zero energy corresponds to the edge states. The lower panels show 
TB-DOS as a function of the wave number at the bottom of the band (left) 
and above the Fermi level (right). The dashed line shows the analytical 
result of Eq. (\ref{levels2}). The wave numbers are: $Q=\sqrt{\epsilon+3}$ 
and $q=\epsilon$, where $\epsilon$ is the energy in units of $t=1$. The 
Gaussian widths have been adjusted in the lower panels in order to show the 
individual energy levels.}
\label{tridos}
\end{figure}

In a triangular cavity with hard walls and a constant potential inside, the 
two-dimensional Schr\"odinger equation has an exact solution
\cite{krishnamurthy1982} with energy levels 
\begin{equation}
\epsilon_{n,m}=\epsilon_0(n^2+m^2-nm),
\label{levels1}
\end{equation}
where $\epsilon_0$ depends on the particle mass and the size of the 
cavity, and $m$ and $n$ are integers with $n\ge 2m \ge 1$. Figure 
\ref{tridos} shows the density of states calculated with the TB model 
(TB-DOS) for a large graphene triangle of 44097 atoms which has a zigzag 
edge. In such a large triangle, DOS is similar to that of an infinite 
graphene sheet except for the appearance of the edge states which appear as 
a sharp peak at zero energy. A detailed study of the energy levels at the 
bottom of the valence band reveals that the level structure is nearly 
exactly described with the analytical formula of Eq. (\ref{levels1}). 
TB-DOS and Eq. (\ref{levels1}) produce the same curve shown in the lower 
left corner of the figure. Note that TB-DOS is plotted here as a function 
of the wave number defined as $Q=\sqrt{\epsilon+3t}$ (The bottom of the 
band is $-3t$). The regular oscillation as a function of $Q$ corresponds to 
the shell structure and the peak amplitude variation (breathing) marks the 
supershell structure\cite{brack1997}. Triangles with an armchair edge show 
a similar supershell structure at the bottom of the band\cite{akola2008}.

At the bottom of the conduction band, close to $\epsilon=0$, the dispersion 
relation of the electron energy is linear, $\epsilon(k)=c \hbar k$, where 
$c$ is the electron velocity. This means that the electrons behave as 
massless particles. If we consider the conduction electrons as free 
particles, we cannot solve the energy eigenvalues from the Schr\"odinger 
equation but should use the relativistic Dirac equation 
\cite{castroneto2008}. However, we choose here a simpler approach and make 
an ansatz that the energy eigenvalues are solutions of the Klein-Gordon 
wave equation with positive energy eigenvalues. This immediately gives 
\begin{equation}
\epsilon_{n,m}=\epsilon_1\sqrt{n^2+m^2-nm)},
\label{levels2}
\end{equation}
which is the same as for normal electrons apart of the square root
dependence of the quantum numbers and a different prefactor. The numerical 
solutions of the TB problem for triangles with a zigzag edge, indeed, show 
that the energy eigenvalues become more-and-more accurately described with 
those of Eq. (\ref{levels2}) when the triangle size increases. Figure 
\ref{tridos} (lower right panel) compares the result of the ansatz of Eq. 
(\ref{levels2}) with the full TB calculation for the large triangle. The 
agreement is nearly perfect up to $\epsilon=0.1$ (corresponding to 0.25 
eV), and the discrepancy at larger energies is due to the increasing 
nonlinearity and anisotropy of the energy bands.

The results suggests that the supershell structure of the triangular
cavity appears in zigzag-edged triangular graphene flakes, but the 
flake should have at least ca. 40000 atoms (i.e., $L\ge40$ nm) before the 
first supershell oscillation becomes clearly visible. We remark that in the 
case of armchair edge, Eq. (\ref{levels2}) is still valid, but also indices 
with $m=n$ are allowed\cite{akola2008}. More detailed results for smaller 
triangles are described in Ref. \cite{heiskanen2008}, where we also show 
that the shell structure is quite robust against edge roughness in the 
close vicinity of the Fermi level.

\section{Shell structure in circular graphene flakes}

\begin{figure}[h]
\includegraphics[width=\columnwidth]{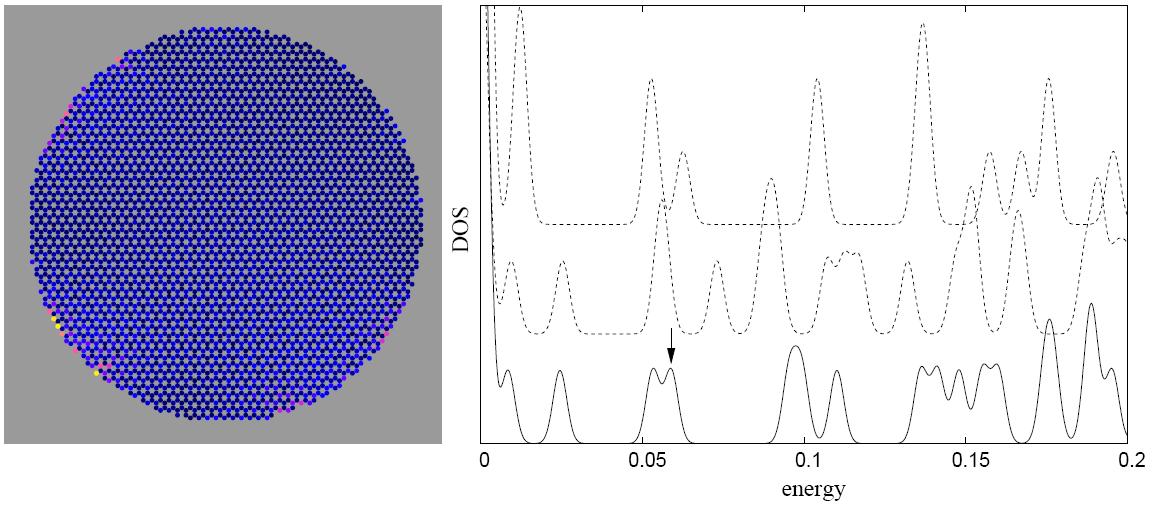}
\caption{Geometry of a circular dot cut out from a graphene sheet (left)
showing electrons density of the state shown with an arrow (black is zero
density, yellow high density).
The three curves (right) show TB-DOS just above the Fermi energy 
for three circular flakes with 3868 (solid line), 3864 (dashed line), and 
3868 atoms (dotted line), respectively, demonstrating the sensitivity of 
the level structure on edge geometry.
}
\label{circledos}
\end{figure}

In the case of a two-dimensional cavity with circular symmetry, the energy 
levels of the Schr\"odinger equation are determined by the zeroes of the 
Bessel functions $B_j$ with integer values $j$. The TB model gives 
corresponding results at the bottom of the valence band, because the 
electrons are well represented by nearly free electrons. Following the 
ideas presented for the triangles one would expect that the energy levels 
close to the bottom of the conduction band could be determined similarly. 
However, in this case, the detailed geometry of the edge (perimeter) has a 
dominant role in determining the energy levels above the Fermi level, and 
the energy spectrum is very sensitive to the number of atoms in the 
circular dot, as shown in figure \ref{circledos}. The circular flakes have 
been obtained by cutting a circle out of an infinite graphene sheet. Note, 
that the actual edge geometry depends not only on the radius but also on 
the site of the center. 

The reason for the size-sensitivity can be traced back to the edge states 
which are present in graphene constructions with zigzag edges. In circular 
dots, the perimeter has short regions of zigzag segments that are mixed 
with other motifs (especially armchair). This roughness causes edge states 
with different energies, which is in sharp contrast to the zigzag triangles 
where all the edge states have exactly zero energy in the TB model. 
Figure \ref{circledos} shows the electron density of one such state
with density maximas at the surface.

Usually, the shell structure is determined by the overall shape of the 
confining potential in metallic and semiconductor quantum dots, and the 
detailed atomic structure does not play any role due to the fact that the 
electron wave length is much larger than the interatomic spacing. This is 
not the case in the circular graphene flakes. Although the wavelenght of 
the "Dirac electron" is still much larger than the interatomic spacing, the 
tendency for localization of electrons close to the zigzag edges destroys 
the simple shell structure, and different circular flakes result in 
qualitatively different electron levels as shown in Fig. \ref{circledos}.

\section{Quantum dots prepared with external potential}

So far, we have studied free graphene flakes where the electron confinement
is determined by the flake edges. In semiconductor heterostructures, 
quantum dots are usually prepared by confining the delocalized conduction 
electrons in a small region with external gates (for a review see 
\cite{reimann2002}). It is expected that a similar technique can be applied 
in the future also for the two-dimensional gas of "Dirac electrons" in 
graphene. External gates form nearly harmonic confinement close to the 
center of the quantum dot. Another possibility for supported graphene 
flakes could be to modify the atomic structure of the substrate so that 
different regions would comprise different elements, and, consequently, 
cause different interaction with the adsorbate. In this case, the 
resulting potential well could be more of the square-well-type than harmonic.

\begin{figure}[h]
\begin{center}\includegraphics[width=0.75\columnwidth]{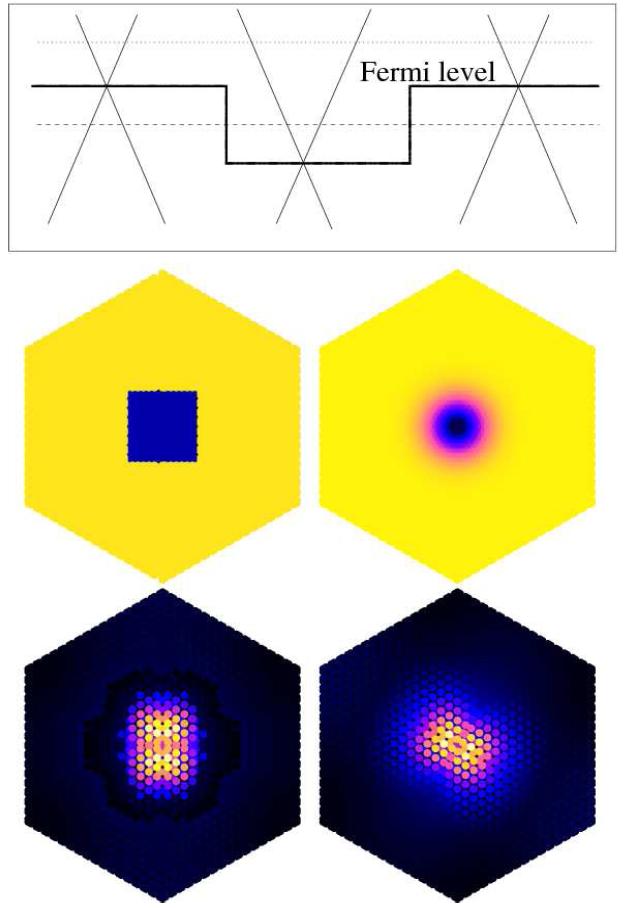}\end{center}
\caption{The upper panel shows an external square well potential in 
graphene and schematically the linear energy bands in different regions. 
Electrons cannot be localized by the potential, because a conduction 
electron inside the well (dashed line) can move out as a valence electron. 
The two figures in the middle display a square-shaped and circular Gaussian 
potential wells in a hexagonal graphene flake. The lower two figures 
illustrate the electron densities of a resonance state above the Fermi 
level (dotted line in the upper panel).
}
\label{wells}
\end{figure}

The situation is different for an external confinement (infinite graphene 
sheet) than for a finite flake with edges. Since there is no band gap in 
graphene, an external potential well cannot bind an electron as 
demonstrated in Fig. \ref{wells}. In addition, this differs considerably 
from the quantum dots manufactured from semiconductor heterostructures, 
where bound electronic states can exist inside the band gap of the 
semiconductor in question.

We have studied the effect of an external confinement by using a large 
hexagonal graphene flake with 4902 atoms. An external potential was added 
at the center of the flake. We considered three different external 
potentials: A circular well, a square-shaped well, and a smooth Gaussian 
potential with circular symmetry. Surprisingly, the results are 
qualitatively similar irrespective of the type of the attractive potential. 
Bound states appear at the bottom of the valence band where the electrons 
act as normal free electrons. In the more interesting region close to the 
Fermi level, no bound states can be observed. However, above the Fermi 
level all the potentials result in strong resonances with a large 
enhancement of the wave function amplitude within the potential well 
region. Figure \ref{wells} shows the densities of the wave functions for 
two such resonances. The wave functions do not decay to zero outside the 
potential well but reach a small and uniform amplitude that goes all the 
way to the flake edge (the small amplitude is not visible Figure 
\ref{wells}).

For a potential barrier, the penetration of a wave function inside an 
apparently forbidden region is often referred to as the Klein 
paradox\cite{castroneto2008}, which has its origin in the Dirac theory of 
massless fermions. As Figure \ref{wells} shows for the TB (band structure) 
model, the wave function penetration inside the ``forbidden region'' is a 
natural consequence of the missing band gap: An electron that appears on 
the conduction band on one side of the step continues as a valence electron 
on the other side.

\section{Conclusions} 

We have studied the possibility of observing electronic shell and 
supershell structure in free graphene flakes. For this purpose, we have
used a simple tight-binding model with one electron per atomic site 
($p_z$ electrons). Despite its simplicity, the TB model describes the
key features of the graphene band structure close to the Fermi points.

In large triangular flakes with zigzag edges, the shell structure of the 
``Dirac electrons'' in the conduction band is the same as for free 
electrons in a triangular cavity. The analytical expression gives the 
energy levels accurately up to $\sim$0.25 eV above the Fermi energy, and 
the number of shells within this region depends on the number of atoms in 
the triangle. A triangle of ca. 40000 atoms ($L\ge40$ nm) shows already 
the first supershell oscillation. 

In the case of circular graphene flakes, the shell structure above the 
Fermi level is dominated by the states that are localized close to the 
zigzag regions of the edges. This makes the shell structure very sensitive,
not only to the radius of the circular flake (number of atoms) but also to 
the location of the center.

Potential wells which are created on an infinite graphene sheet with 
external potentials (e.g. external gates, inhomogeneous substrate) cannot 
localize electrons. This is a consequence of the missing band gap in the 
graphene band structure. However, such potential wells cause resonance 
states above the Fermi level, which can strongly affect the conductance of 
narrow graphene strips.

\vskip0.3truecm

{\bf Acknowledgments}
This work has been supported by the Academy of Finland.

\end{document}